%% file: arxiv.tex
\documentclass[letterpaper,twocolumn,10pt]{article}
\usepackage{usenix}

\usepackage{amsmath,amssymb,amsfonts}
\usepackage{graphicx}
\usepackage{booktabs}
\usepackage{multirow}
\usepackage{array}
\usepackage{colortbl}
\usepackage{balance}
\usepackage{rotating}
\usepackage{xcolor}
\usepackage{enumitem} 
\usepackage{wrapfig}
\usepackage{xurl}
\usepackage{tikz}
\usepackage{pgfplots}
\pgfplotsset{compat=1.18}
\usetikzlibrary{shapes.geometric,arrows.meta,positioning,fit,backgrounds,calc,shadows}
\usepackage{subcaption}
\usepackage{siunitx}
\usepackage{mathtools}
\usepackage{bm}
\usepackage[most]{tcolorbox}
\usepackage{pifont}
\usepackage{lineno}
\usepackage{doi}

\newcommand{\cmark}{\ding{51}}
\newcommand{\xmark}{\ding{55}}

\definecolor{vulncol}{RGB}{214,39,40}
\definecolor{fixcol}{RGB}{44,160,44}
\definecolor{partialcol}{RGB}{255,127,14}
\definecolor{netcol}{RGB}{31,119,180}
\definecolor{tableheader}{RGB}{240,240,240}
\definecolor{findingbg}{RGB}{255,250,230}
\definecolor{findingframe}{RGB}{200,160,40}
\definecolor{authorburgundy}{HTML}{800020}
\usepackage{tikz}
\usetikzlibrary{arrows.meta,positioning,fit,calc}

\hypersetup{
	colorlinks=true,
	linkcolor=blue!60!black,
	citecolor=green!50!black,
	urlcolor=purple!70!black,
	pdfstartview={FitH}
}

\newtcolorbox{findingbox}{
	colback=findingbg,
	colframe=findingframe,
	boxrule=0.6pt,
	arc=2pt,
	left=4pt, right=4pt, top=3pt, bottom=3pt,
	fonttitle=\bfseries\small
}

\begin{document}

    \title{The Illusion of Local Privacy:
Confidentiality Boundary Failures in Consumer LLM Serving Systems}

     \author{
     	{\rm Youssef~Hamdi~Zafan~Ibrahim} \\
     	\textit{Independent Researcher} \\
     	{\tt youssefhamdi329@gmail.com} \\
     	{\scriptsize ORCID: \href{https://orcid.org/0009-0000-9563-8020}{\textcolor{authorburgundy}{0009-0000-9563-8020}}} \\
     	{\scriptsize Corresponding author}
     	\and
        {\rm Muhammad~Ikram} \\
     	\textit{Macquarie University} \\
     	{\tt muhammad.ikram@mq.edu.au} \\
     	{\scriptsize ORCID: \href{https://orcid.org/0000-0003-2113-3390}{\textcolor{authorburgundy}{0000-0003-2113-3390}}}
        \and
     	{\rm Mohammed~Khalaf~Salama} \\
     	\textit{Independent Researcher} \\
     	{\tt mkhalafsalama@gmail.com} \\
     	{\scriptsize ORCID: \href{https://orcid.org/0009-0007-6508-1437}{\textcolor{authorburgundy}{0009-0007-6508-1437}}}
     }

	\maketitle
	\begin{abstract}

Running large language models (LLMs) locally is often considered more private than cloud-hosted inference because user prompts remain on the device. We ask whether keeping inference local is, by itself, sufficient to keep those prompts confidential. Our results show that it is not: prompt confidentiality also depends on how the surrounding serving software handles prompt data before, during, and after inference.

We systematically examine prompt confidentiality across consumer local-LLM serving systems. Instead of treating the local deployment as a single trusted environment, we examine four boundaries at which prompt confidentiality can fail: model loading, runtime memory, wrapper-level persistence, and the serving interface. To study these boundaries, we develop \textbf{LLAnalyzer}, a measurement framework that lets us test each boundary separately and trace observed failures to the software component responsible.

We apply LLAnalyzer to four open-weight model families and two consumer deployment platforms and find markedly different behaviour across these boundaries. In a 24-hour AFL++ campaign comprising more than $1.2\times10^7$ executions, we observe no parser crashes or successful malformed GGUF loads within the explored state space. Runtime memory tells a different story: we recover prompts after inference because multiple plaintext representations survive in allocator-managed memory, and sanitisation reduces this residue without eliminating it. We also find that consumer wrappers can extend prompt lifetime through plaintext persistence. At the serving boundary, we uncover a previously undocumented authorization flaw in \texttt{llama.cpp} that allows one authenticated client to restore another tenant's saved conversation state; the attack succeeds in 200/200 controlled trials (Wilson 95\% CI: $[98.1\%, 100.0\%]$). Separately, we show that shared prompt-prefix caching exposes a remote timing oracle that remains fully distinguishable under both loopback and real wide-area network conditions (AUC $=1.000$ for each of the four evaluated models; mean RTT 78.4\,ms), and that separation persists up to broadband WAN jitter ($\sigma\le 100$\,ms) in a Monte Carlo sensitivity sweep.

Our measurements show why we should distinguish \emph{local execution} from \emph{prompt confidentiality}. Keeping inference on-device removes one source of exposure, but it does not control what the local serving stack subsequently does with prompt data. We therefore argue that local LLM systems need explicit guarantees for prompt lifetime, persistent storage, and tenant isolation in addition to localising computation.

\end{abstract}
	
	\vspace{0.5em}\noindent\textbf{Index Terms:}
	Large language models, local inference, privacy, memory forensics,
	authorisation, KV-cache, side channels, responsible disclosure, GGUF,
	\texttt{llama.cpp}.






\section{Introduction}
\label{sec:introduction}

Large language models (LLMs) can now run entirely on personal computers through tools such as \emph{LM Studio}~\cite{lmstudio}, \emph{Ollama}~\cite{ollama}, and \texttt{llama.cpp}~\cite{llamacpp}. For privacy-sensitive workloads, the appeal is straightforward: prompts can be processed without sending them to a cloud inference provider. But what does ``local'' actually guarantee about a prompt after it reaches the serving software?

We find that this distinction matters. Keeping computation on-device determines where inference takes place; it says much less about what happens to a prompt while inference is being prepared, executed, and torn down. A single request may pass through model-loading code, HTTP and JSON parsers, chat-template processing, runtime memory, application storage, and a serving interface shared by multiple clients. These components make independent decisions about copying, retaining, persisting, and exposing prompt data. We therefore ask whether the privacy intuition associated with local inference survives this broader software lifecycle.

There are reasons to examine this question now. Internet measurements have reported publicly reachable local-LLM deployments ~\cite{shodan_query2026,censys_ollama_drama2026,thn_ollama175k2026}, alongside reports of exposed serving interfaces and prompt-related information~\cite{upguard_llamacpp2025,opendoors_lmstudio2026}. Research on LLM privacy, meanwhile, has largely examined different parts of the problem: memorisation of training examples ~\cite{carlini2023memorization,carlini2025}, timing leakage from shared KV-cache optimisations~\cite{wu2025iknowwhatyouasked,safekv2025,prefixwall2026}, and reconstruction attacks against cached model state ~\cite{shadowcache2026,optileak2026}. Our own passive snapshot (4 July 2026) identifies 32{,}723 publicly reachable Ollama endpoints and 1{,}221 \texttt{llama.cpp} endpoints, 854 of which expose complete model metadata, establishing that the exposure is not confined to laboratory deployments. Much less is known about what happens to a user's prompt across the ordinary software stack of a consumer local-LLM deployment.

This leads us to a systems question: \emph{when inference stays on the user's machine, which components of the serving stack actually determine whether the prompt stays confidential?} Answering it requires more than searching for individual implementation bugs. If we treat the entire local stack as one trusted component, a failure observed at the serving interface is indistinguishable, conceptually, from one caused by runtime-memory retention or application-level persistence. It also becomes difficult to determine where a mitigation should be applied.

We instead study prompt confidentiality as a composition of explicit software boundaries. We develop \textbf{LLAnalyzer}, a measurement framework that separates a consumer local-LLM serving stack into four boundaries that we can exercise independently: \emph{(i)} an \textbf{Integrity Boundary}, covering model admission and loading; \emph{(ii)} a \textbf{Lifetime Boundary}, covering the lifetime of prompt data in runtime memory; \emph{(iii)} a \textbf{Persistence Boundary}, covering prompt material retained by wrapper applications; and \emph{(iv)} an \textbf{Isolation Boundary}, covering separation between clients of the serving interface. For each boundary, we inject controlled inputs, acquire boundary-specific evidence, and trace an observed failure back to the software layer in which it occurs.

Applying LLAnalyzer to four open-weight model families and two consumer deployment platforms produces a markedly uneven security picture. At the integrity boundary, a 24-hour AFL++ campaign executes more than $1.2\times10^{7}$ parser tests without producing a parser crash, sanitizer violation, or successful malformed GGUF~\cite{gguf} load within the explored state space. The runtime boundary behaves differently. We recover prompt plaintext after inference and trace the residue to multiple representations created during ordinary request processing and subsequently retained in allocator-managed memory. Sanitisation reduces this residue substantially, but does not eliminate historical prompt recovery.

The failures are not confined to volatile memory. We find that consumer wrappers can extend prompt lifetime through plaintext persistence. At the serving interface, we identify a previously undocumented authorization flaw in \texttt{llama.cpp}: an authenticated client can restore conversation state belonging to another tenant. The operation succeeds in all 200 controlled trials across the four evaluated model families (Wilson 95\% CI: $[98.1\%, 100.0\%]$). Separately, we measure a remote timing signal introduced by shared prompt-prefix caching and find that cached and uncached prefixes remain fully distinguishable under both loopback and real wide-area network conditions (AUC $=1.000$ for each model), with separation persisting up to broadband WAN jitter ($\sigma\le 100$\,ms).

These results change how we think about the privacy claim attached to local inference. Moving computation from a cloud provider onto a user's machine removes an important external trust relationship, but it does not establish what happens to prompt data inside the local serving stack. In the systems we study, confidentiality depends on the composition of model admission, runtime data lifetime, persistent storage, and client isolation. Local execution is therefore one condition in this trust model, rather than its endpoint.

\textbf{Contributions.}
Our work makes the following contributions:

\begin{itemize}

\item \textbf{A boundary-oriented model of local-LLM prompt
confidentiality.}
We formulate prompt confidentiality as a systems property spanning model admission, runtime lifetime, application persistence, and client isolation. This decomposition gives us explicit security objectives against which each layer can be measured.

\item \textbf{LLAnalyzer.}
We design and implement LLAnalyzer, a measurement framework for exercising these boundaries independently and attributing observed prompt-confidentiality failures to the software components responsible for them.

\item \textbf{An empirical study of confidentiality across the local
serving stack.}
Across four model families and two consumer deployment platforms, we find sharply different behaviour across boundaries: no observed parser failure within our fuzzing campaign, persistent post-inference prompt residue, wrapper-level plaintext persistence, and a previously undocumented cross-tenant authorization flaw. We additionally characterize a remote timing signal arising from prompt-prefix reuse, including real wide-area validation and a Monte Carlo jitter sweep.

\item \textbf{Mechanisms, mitigations, and disclosure.}
We trace the mechanisms underlying the observed failures, evaluate runtime sanitisation and its performance cost, and responsibly disclose the identified vulnerabilities to the affected maintainers. We release LLAnalyzer and the supporting experimental artifacts to enable reproduction and further measurement.
\end{itemize}

\noindent\textbf{Paper Structure.}
Section~\ref{sec:framework} defines our confidentiality boundaries and threat model. Section~\ref{sec:methodology} describes LLAnalyzer and our measurement methodology. Section~\ref{sec:evaluation} evaluates each boundary and investigates the mechanisms behind the observed failures. In Section~\ref{sec:related}, we position our findings against prior work, and conclude our work in Section~\ref{sec:conclusion}. 

\begin{figure*}[ht]
	\centering
	\includegraphics[width=0.90\textwidth]{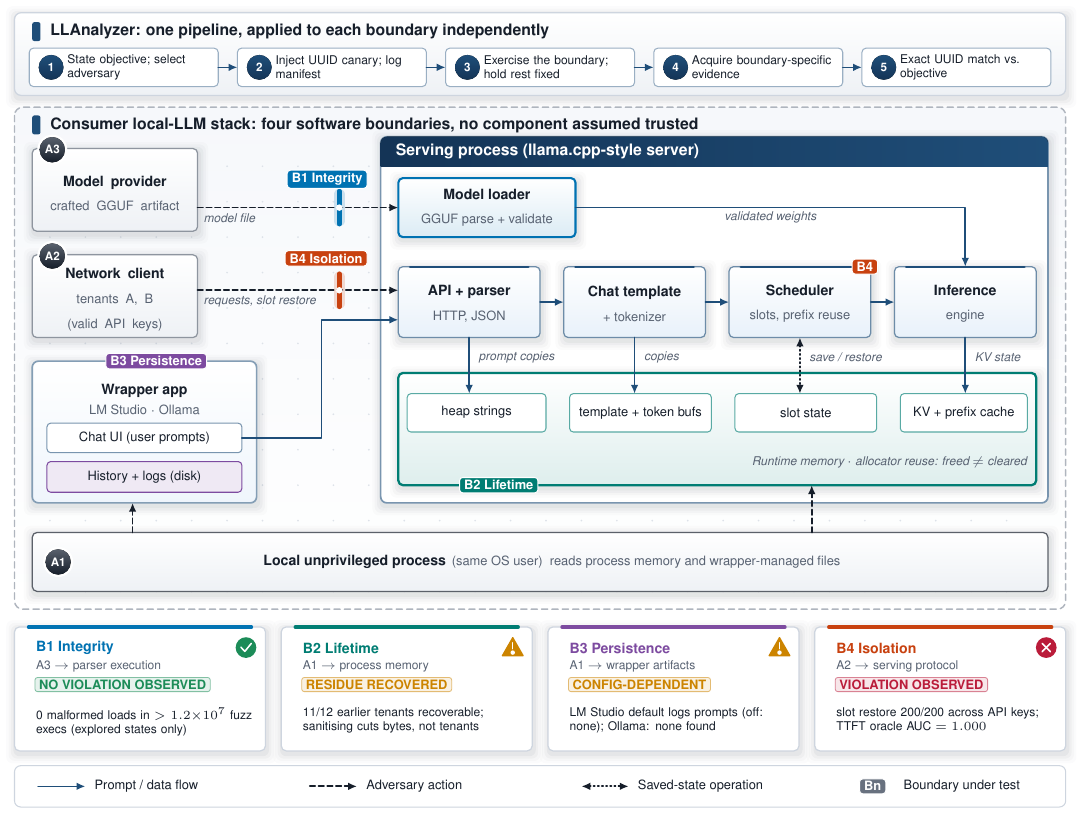}
	\caption{Architecture and evidence map for LLAnalyzer. The concrete
		request path is shown above the four confidentiality boundaries:
		model-file integrity (B1), runtime lifetime (B2), wrapper-managed
		persistence (B3), and shared serving-state isolation (B4).}
	\label{fig:pb-llanalyzer}
\end{figure*}

\begin{figure}[!t]
	\centering
	\includegraphics[width=0.88\columnwidth]{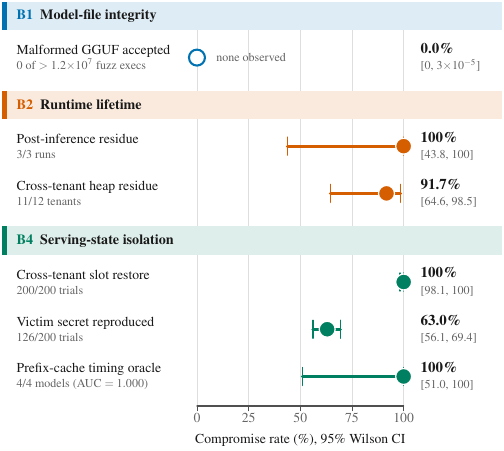}
	\caption{Boundary-level compromise rates with Wilson 95\% confidence intervals. Each row reports the observed compromise rate for a single experiment in Section~\ref{sec:evaluation}, with the point estimate shown as a filled marker and the Wilson interval as a horizontal line with end caps. B1--B4 refer to the boundaries defined in Figure~\ref{fig:pb-llanalyzer}.}
	\label{fig:forest}
\end{figure}

\section{Confidentiality Boundary Framework}
\label{sec:framework}

We begin by defining what prompt confidentiality means in a consumer
local-LLM deployment. We do not treat the deployment as a single trusted
environment. Instead, we follow a prompt through the serving stack and
identify the points at which different software components assume
responsibility for protecting it. This gives us four boundaries with
distinct security objectives and, importantly, distinct ways of failing.

\subsection{Prompt Lifecycle in Consumer LLM Serving Systems}

A typical local-LLM request begins in a consumer application such as
\emph{LM Studio} or \emph{Ollama} and eventually reaches an inference
engine such as \texttt{llama.cpp}. Between receiving the user's text and
returning a response, the stack loads the model, parses and transforms
the request, constructs the prompt, tokenizes it, manages inference
state, and may record conversation state or expose it through a serving
API.

Although this appears to the user as one local operation, the prompt does
not remain in one representation or under the control of one component.
It can be copied during parsing, transformed by a chat template, retained
in runtime memory, written to application storage, or associated with
state exposed through a serving interface. A confidentiality guarantee
at one of these stages therefore says little about the others.

Figure~\ref{fig:pb-llanalyzer} follows this lifecycle and shows where
LLAnalyzer places the corresponding confidentiality boundaries.

\subsection{Confidentiality Boundaries}

We define four boundaries according to the security decision being made
about prompt data.

\noindent\textbf{Integrity Boundary.}
Before inference begins, the model-loading pipeline decides whether an
input model artifact can safely enter execution. GGUF parsing, metadata
validation, tensor validation, and model initialisation form this
boundary. Its objective is to prevent malformed model artifacts from
compromising trusted model loading or subsequent execution.

\noindent\textbf{Lifetime Boundary.}
Once a request is processed, prompt data can exist simultaneously in
several runtime representations produced by parsing, template
construction, tokenisation, request processing, and memory allocation.
The lifetime boundary concerns whether those representations remain
recoverable after they are no longer required for inference.

\noindent\textbf{Persistence Boundary.}
Outside process memory, wrapper applications may retain conversation
histories, logs, caches, temporary files, and other application
artifacts. This boundary concerns whether prompt data are written to
persistent storage, for how long, and under what configuration.

\noindent\textbf{Isolation Boundary.}
When multiple clients interact with the same serving process, the
serving interface must keep their state separate. Authentication,
authorization, conversation-state management, cache sharing, and session
handling form this boundary. Its objective is to prevent one client from
learning prompt or conversation information belonging to another.

These boundaries are deliberately separated because satisfying one does
not imply satisfying another. A model may load safely while prompts
remain in memory; runtime memory may be sanitised while a wrapper writes
the same prompt to disk; and a locally protected process may still expose
another tenant's state through its API. We use this decomposition to
identify which guarantee is being tested in each experiment.

\subsection{Threat Model}
\label{sec:threat}

We consider adversaries that can exercise one of these boundaries without
already controlling the operating system or inference process. We assume
that the underlying operating system, hardware, and cryptographic
primitives operate correctly unless an experiment explicitly states
otherwise. Our goal is therefore not to model a fully compromised host,
but to determine what prompt information becomes available through the
ordinary privileges and interfaces surrounding a local-LLM deployment.

\noindent\textbf{${A}_1$ (Local Process).}
${A}_1$ is an unprivileged process running under the same operating-system
user account as the local-LLM stack. This captures desktop environments
in which browser extensions, IDE plugins, utilities, and other user
applications coexist with the inference server. We use ${A}_1$ to examine
whether prompt data remain recoverable in runtime memory or application
artifacts beyond their intended lifetime.

\noindent\textbf{${A}_2$ (Network Client).}
${A}_2$ is an authenticated client restricted to the published serving
interface. It has no direct access to process memory, the host filesystem,
or the underlying operating system. We use this adversary to test whether
the serving protocol preserves isolation between clients.

\noindent\textbf{${A}_3$ (Malicious Model Provider).}
${A}_3$ can supply a crafted or malformed GGUF artifact to an otherwise
unmodified inference engine. We use this capability to test whether the
model-loading pipeline rejects malformed artifacts without compromising
trusted execution.

\begin{table}[!t]
\centering
\small
\caption{Confidentiality boundaries and corresponding adversary models.}
\label{tab:threatmodel}
\begin{tabular}{p{1.2cm}p{0.8cm}p{5.2cm}}
\toprule
\textbf{Boundary} & \textbf{Adv.} & \textbf{Security objective} \\
\midrule
Integrity    & ${A}_3$ & Reject malformed model artifacts without
                         compromising trusted execution. \\
Lifetime     & ${A}_1$ & Minimize recoverable prompt plaintext after
                         inference. \\
Persistence  & ${A}_1$ & Prevent unintended prompt retention on local
                         storage. \\
Isolation    & ${A}_2$ & Prevent one client from recovering another
                         client's prompt or conversation state. \\
\bottomrule
\end{tabular}
\end{table}

Table~\ref{tab:threatmodel} connects each boundary to the adversary capable
of challenging it and to the property we subsequently measure. This
mapping also determines what constitutes a confidentiality failure in
each experiment.

\section{LLAnalyzer}
\label{sec:methodology}

LLAnalyzer operationalizes the four boundaries above as controlled
measurement experiments; to our knowledge, it is the first framework to
measure prompt confidentiality end-to-end across this stack. Its
central design principle is simple: when
testing one boundary, we vary the mechanism relevant to that boundary
while keeping the remainder of the serving configuration fixed. The
result is not a single vulnerability scanner, but a collection of
boundary-specific measurement harnesses connected by a common
experimental workflow.

\subsection{Design Goals}

We designed LLAnalyzer around three goals.

\noindent{\bf Boundary attribution.}
A recovered prompt is useful evidence only if we can determine how it
became exposed. LLAnalyzer therefore associates every measurement with a
specific boundary and collects evidence from the software component that
implements that boundary. This lets us distinguish, for example,
runtime-memory residue from wrapper persistence or protocol-level
cross-tenant disclosure.

\noindent{\bf Methodological consistency.}
The boundaries cannot all be tested using the same instrumentation:
model admission requires parser testing, runtime lifetime requires memory
forensics, persistent storage requires artifact analysis, and tenant
isolation requires protocol-level experiments. We nevertheless impose
the same experimental sequence on each measurement so that the resulting
evidence answers the same question: did the target boundary satisfy its
security objective?

\noindent{\bf Reproducibility.}
We establish explicit experimental ground truth using randomly generated
UUID canaries, fixed software versions, byte-identical model artifacts,
and controlled execution configurations. The implementation, experimental
configurations, and supporting artifacts are released as described in the ``Open Science" section (see
page \# 14). 
For each experiment, LLAnalyzer generates
a fresh canary and embeds it in the submitted prompt, providing an
experiment-specific marker whose subsequent recovery can be attributed
to the originating request rather than inferred from surrounding content.

\subsection{Framework Architecture}

Figure~\ref{fig:pb-llanalyzer} shows how these measurement components map
onto the serving stack. For each boundary, LLAnalyzer follows five steps:
(\textit{i}) state the security objective;
(\textit{ii}) select the corresponding adversary;
(\textit{iii}) exercise the target boundary under a controlled
configuration;
(\textit{iv}) acquire evidence from that boundary; and
(\textit{v}) test the evidence against the security objective.

The instrumentation used in Step~(\textit{iv}) changes with the boundary;
the experimental logic does not. This gives us a common structure for
interpreting evidence obtained from otherwise different parts of the
serving system.

\subsection{Boundary-Oriented Measurement}

\begin{table}[!t]
\centering
\small
\caption{Boundary-specific evidence collected by LLAnalyzer.}
\label{tab:measurement_pipeline}
\begin{tabular}{p{1.6cm}p{2.2cm}p{3.1cm}}
\toprule
\textbf{Boundary} & \textbf{Evidence} & \textbf{Measured property} \\
\midrule
Integrity &
Parser execution &
Safe rejection of malformed artifacts \\

Lifetime &
Process memory &
Post-inference prompt recoverability \\

Persistence &
Wrapper artifacts &
Prompt retention on persistent storage \\

Isolation &
Serving protocol &
Cross-client information separation \\
\bottomrule
\end{tabular}
\end{table}

Table~\ref{tab:measurement_pipeline} shows the evidence source used for
each boundary. During an experiment, we change only the conditions needed
to exercise the target boundary and preserve the remaining configuration.
Where a UUID canary is subsequently recovered, LLAnalyzer uses the
boundary-specific evidence to determine the software path through which
that disclosure occurred.
\subsection{Measurement Pipeline}

Although each confidentiality boundary requires different
instrumentation, every LLAnalyzer experiment follows the same measurement
pipeline, shown in Figure~\ref{fig:pb-llanalyzer}.

\textit{First, establish ground truth.}
For each inference request $r$, LLAnalyzer generates a fresh UUIDv4 using
the operating system's cryptographically secure random-number generator
and embeds its canonical representation directly into the submitted
prompt as \texttt{CANARY:<uuid>}. Before inference begins, the harness
records
$\langle
r,\mathrm{UUID}_r,\mathrm{boundary},\mathrm{timestamp}
\rangle$
in the experiment manifest. This provides a deterministic mapping between
every injected marker and its originating request. A UUIDv4 contains 122 random bits, making an accidental exact match
negligibly probable. Before each experiment, LLAnalyzer additionally
checks that the newly generated canary is absent from the target evidence
source. An exact post-inference match can therefore be associated with
the corresponding injected request with negligible probability of
accidental collision.

\textit{Second, exercise the target boundary.}
LLAnalyzer executes the experiment under a controlled configuration,
isolating the boundary under evaluation while holding unrelated
experimental factors constant wherever possible. The specific condition
being exercised depends on the corresponding security objective.

\textit{Third, acquire boundary-specific evidence.}
The evidence source depends on the boundary: parser execution for the
Integrity Boundary, process memory for the Lifetime Boundary,
wrapper-managed artifacts for the Persistence Boundary, and
protocol-level observations for the Isolation Boundary.

\textit{Finally, test the boundary objective.}
LLAnalyzer searches the acquired evidence for exact UUID matches and maps
each recovered occurrence to its originating request through the
experiment manifest. A boundary failure is recorded when recovery
satisfies the predefined disclosure criterion for that boundary. The
precise success and failure criteria are specified with the corresponding
experiments in Section~\ref{sec:evaluation}.

\subsection{Experimental Configuration}

We conduct the experiments in controlled Linux environments using fixed
hardware, software revisions, inference parameters, and model artifacts
unless an experiment explicitly requires otherwise. Experiments involving
GPU execution are additionally repeated under native Ubuntu to distinguish
the measured behaviour from artifacts specific to the WSL2 environment.

We select four open-weight model families spanning different parameter
scales and architectural designs. All evaluated artifacts use GGUF
Q4\_K\_M quantisation and were obtained from the
\texttt{lmstudio-community} organization on Hugging Face. Before each
experimental campaign, LLAnalyzer records and verifies the SHA-256 digest
of every model artifact to ensure that identical model bytes are used
across experimental conditions. Table~\ref{tab:models} summarizes the
evaluated models.

\begin{table}[!th]
    \centering
    \resizebox{\columnwidth}{!}{%
        \footnotesize
        \begin{tabular}{@{}llll@{}}
            \toprule
            Model & Params & Architecture & Size (Bytes) \\
            \midrule
            NVIDIA Nemotron-3-Nano
                & 4\,B & hybrid attn/SSM & 2{,}837{,}072{,}896 \\
            Qwen3.5
                & 9\,B & transformer & 5{,}627{,}044{,}256 \\
            Gemma-4-E4B-it
                & 7.5\,B & transformer & 5{,}335{,}289{,}664 \\
            Phi-4-reasoning-plus
                & 14\,B & transformer (reasoning) & 9{,}053{,}116{,}128 \\
            \bottomrule
        \end{tabular}%
    }
    \caption{Open-weight models evaluated by LLAnalyzer. All model 
    artifacts use GGUF Q4\_K\_M quantisation and were obtained from the 
    \texttt{lmstudio-community} organization on Hugging Face.}
    \label{tab:models}
\end{table}

\subsection{Experimental Controls and Statistical Analysis}

We use controlled comparisons to separate boundary-specific effects from
changes elsewhere in the serving stack. Within each experiment, hardware,
operating system, model artifact, software revision, and inference
parameters remain fixed unless the corresponding variable is itself
under evaluation. When comparing configurations, we use identical prompt
inputs and model artifacts wherever applicable.

For each boundary, we define an explicit success and failure criterion
and apply it consistently across all trials reported for that experiment.
A disclosure is recorded only when an exact UUID match satisfies the
corresponding boundary-specific criterion. For repeated binary experiments, we report the observed proportion with
Wilson 95\% confidence intervals~\cite{wilson1927}. Paired binary
comparisons use McNemar's test~\cite{mcnemar1947} where applicable.
Continuous measurements, including inference latency and recovered-memory
volume, are summarized using descriptive statistics appropriate to the
measurement.




\section{Evaluation}
\label{sec:evaluation}

We use LLAnalyzer to evaluate prompt confidentiality across the four
boundaries defined in Section~\ref{sec:framework}. We organize the
evaluation around research questions rather than individual measurement
tools. For each question, we state the security property under test,
describe the experimental design, present the resulting evidence, and
interpret what the observations establish about the corresponding
boundary.

Our evaluation addresses four research questions:

\begin{itemize}[left=0pt]

\item \textbf{RQ1. Integrity Boundary.}
Can malicious model artifacts bypass model-admission controls or trigger
memory-safety failures before inference begins?

\item \textbf{RQ2. Lifetime Boundary.}
Does prompt plaintext remain recoverable from runtime memory after
inference, and if so, what mechanisms determine its persistence and
mitigation?

\item \textbf{RQ3. Persistence Boundary.}
Do consumer wrapper applications extend prompt lifetime by retaining
prompt data in persistent local artifacts?

\item \textbf{RQ4. Isolation Boundary.}
Can one client learn another client's prompt information through shared
serving state or remotely observable side channels?

\end{itemize}

Together, these questions test our central hypothesis: 
%
\emph{Local execution alone does not establish prompt confidentiality;
confidentiality depends on the security properties enforced across the
software boundaries through which prompt data pass.} 
%
We evaluate the boundaries in prompt-lifecycle order. We begin with model
admission (RQ1), then follow prompt data through runtime memory (RQ2) and
wrapper-managed persistent storage (RQ3), before examining information
separation between clients of a shared serving instance (RQ4). For each
experiment, we isolate the target boundary while holding unrelated
experimental factors constant wherever possible, following the controls
described in Section~\ref{sec:methodology}.

\subsection{RQ1. Integrity Boundary}
\label{sec:rq1}

Before a prompt can be processed, the serving framework must parse and
initialize the supplied GGUF model artifact. A failure at this stage can
invalidate assumptions made by later confidentiality measurements: if a
malformed model compromises the serving process before inference begins,
subsequent prompt disclosures may be consequences of an already
compromised execution environment. We therefore ask: \emph{Can a maliciously crafted GGUF artifact bypass parser validation or trigger memory-safety failures before inference begins?}

{\bf Experimental Design.}
We exercise the model-admission path using two complementary strategies. \textit{First}, we construct structured malformed GGUF artifacts by
systematically modifying valid model files while preserving their overall
format. The mutations target tensor metadata, architecture identifiers,
tensor dimensions, allocation sizes, offsets, RoPE parameters, and
header fields. This approach exercises semantic validation paths rather
than relying solely on arbitrary byte corruption. \textit{Second}, we conduct a 24-hour coverage-guided fuzzing campaign
using AFL++~\cite{fuzzing_taxonomy2024}, with AddressSanitizer (ASan) and
UndefinedBehaviorSanitizer (UBSan) enabled. Starting from valid GGUF
inputs, AFL++ explores additional parser states while we monitor for
crashes, sanitizer violations, memory corruption, and malformed inputs
that progress into model initialization. The structured corpus targets parser states selected from our
understanding of the GGUF format, whereas coverage-guided fuzzing explores
additional states without requiring us to enumerate them in advance.

{\bf Evidence.}
Table~\ref{tab:model_boundary_results} summarizes the results. Every
structured malformed artifact was rejected before model initialization.
We observed no malformed model load, parser crash, abnormal process
termination, or sanitizer violation. The fuzzing campaign executed more than $1.2\times10^{7}$ parser
invocations over 24 hours. ASan and UBSan reported no detected
memory-safety or undefined-behaviour violations during these executions,
and no malformed input progressed beyond parser validation into model
initialization. We observed the same outcome for the tested artifacts
across all four model families.

\begin{table}[!t]
\centering
\caption{Integrity-boundary evaluation using structured malformed GGUF
artifacts and a 24-hour AFL++ coverage-guided fuzzing campaign.}
\label{tab:model_boundary_results} 

\footnotesize
\begin{tabular}{lcc}
\toprule
\textbf{Metric} & \textbf{Observation} & \textbf{Outcome} \\
\midrule
Coverage-guided executions & $>1.2\times10^7$ & Completed \\
Model families             & 4                 & Complete  \\
Sanitizers                 & ASan + UBSan       & Enabled   \\
\midrule
Parser rejection           & 100\%              & \cmark \\
Malformed model loaded     & 0                  & \xmark \\
Parser crashes             & 0                  & \xmark \\
ASan violations            & 0                  & \xmark \\
UBSan violations           & 0                  & \xmark \\
Memory corruption          & 0                  & \xmark \\
\bottomrule
\end{tabular}
\end{table}

{\bf Analysis.}
Within the explored state space, model admission behaves differently from
the downstream boundaries examined later in the evaluation. Our
structured mutations and coverage-guided fuzzing did not identify a path
by which a malformed GGUF artifact bypassed validation, entered model
initialization, or produced a detected memory-safety failure. This result is useful because it bounds the interpretation of the
experiments that follow. Under our evaluated configuration, the
confidentiality failures reported in subsequent sections occur after
successful model admission rather than being preceded by an observed
parser compromise. We can therefore investigate runtime lifetime,
persistent storage, and client isolation without attributing those
observations to a model-admission failure detected by our experiments.

{\bf Scope of the Negative Result.}
Our measurements do not establish that the GGUF parser is universally
secure. Structured mutation necessarily covers only the parser states
represented by the selected mutation classes, and a finite fuzzing
campaign explores only a subset of the possible GGUF input space. We therefore bound this result to the tested artifacts, software
configuration, mutation strategy, and fuzzing campaign. Within that
scope, two complementary approaches produced the same outcome:
semantically targeted malformed artifacts were rejected before
initialization, and more than $1.2\times10^{7}$ coverage-guided
executions under ASan and UBSan produced no observed parser bypass,
sanitizer violation, or malformed model load. Undiscovered parser
vulnerabilities may nevertheless exist outside the explored state space.

{\bf Takeaway.}
Our finding provides no evidence of an Integrity-Boundary violation under
the evaluated conditions. All structured malformed artifacts were
rejected, and the 24-hour coverage-guided campaign produced no observed
parser bypass, sanitizer violation, or malformed model load across more
than $1.2\times10^{7}$ executions. This result establishes the
experimental starting point for the remaining research questions: we
next examine confidentiality failures that arise after successful model
loading, beginning with the lifetime of prompt plaintext in runtime
memory.

\subsection{RQ2. Lifetime Boundary}
\label{sec:rq2}

Having found no Integrity-Boundary violation under the evaluated
conditions in RQ1, we next examine what happens to prompt data after
successful model admission. During inference, a prompt passes through
multiple runtime representations, including formatted templates, JSON
request structures, token buffers, temporary application objects, and
allocator-managed memory. Although these representations are transient
from the application's perspective, their underlying contents may
outlive the request that created them.

This section therefore addresses the following research question: \emph{Does runtime memory preserve prompt confidentiality after
inference, or does prompt plaintext remain recoverable beyond execution
completion?}  We address this question through four progressively deeper
sub-questions. We first establish whether plaintext remains recoverable
after inference (RQ2.1), trace the mechanisms responsible for its
persistence (RQ2.2), evaluate whether runtime sanitisation can reduce
this exposure at acceptable cost (RQ2.3), and finally examine whether
GPU-assisted execution changes these confidentiality properties (RQ2.4).

\subsubsection{RQ2.1: Does Prompt Plaintext Persist After Inference?}
\label{sec:rq2_plaintext}

Our finding identified no parser bypass, sanitizer violation, or malformed
model load under the evaluated conditions. We therefore continue along
the prompt lifecycle and examine runtime state following successful
model admission. Specifically, we ask whether prompt plaintext remains
recoverable from process memory after inference has completed.


{\bf Evidence.} Using the UUID-based forensic methodology described in
Section~\ref{sec:methodology}, LLAnalyzer acquires a complete
post-inference memory snapshot immediately after response generation.
Each prompt contains randomly generated UUID canaries, allowing every
recovered plaintext copy to be attributed to its originating inference
request.

Table~\ref{tab:residue} presents direct evidence of runtime prompt
persistence. Across three independent executions, every measured
inference leaves between 13 and 14 recoverable plaintext prompt copies
after response generation has completed. Although the precise copy
count varies slightly, prompt recovery is reproduced across every
measured execution, indicating that the observation is systematic
rather than specific to a single execution. The practical significance of this residue becomes clearer when the
serving process handles successive users. We therefore execute twelve
sequential tenant sessions within the same serving process and acquire
a single memory snapshot after the sequence completes. Prompt
information from 11 of the 12 tenants remains recoverable, comprising
110 plaintext copies and approximately 10.8\,MB of recovered data (Wilson 95\% CI: $[64.6\%, 98.5\%]$). This result shows that runtime residue is not confined to the most
recent request. Prompt representations associated with earlier tenants
survive subsequent inference requests within the same long-running
process.

\begin{table}[!t]
    \centering
    \caption{Per-tenant heap residue after inference completion. The evaluated implementation consistently leaves 13--14 recoverable plaintext prompt copies per tenant across three independent executions.}
    \label{tab:residue}
    \footnotesize
    \setlength{\tabcolsep}{4pt}
    \begin{tabular}{@{}lcccc@{}}
        \toprule
        Run & Tenant T1 & Tenant T2 & Tenant T3 & Tenant T4 \\
        \midrule
        \texttt{run1} & 13 & 13 & 13 & 13 \\
        \texttt{run2} & 13 & 13 & 13 & 13 \\
        \texttt{run3} & 14 & -- & -- & -- \\
        \bottomrule
    \end{tabular}
\end{table}


{\bf Analysis.} The two experiments establish complementary properties of runtime
residue. First, post-inference memory inspection shows that prompt
plaintext survives response generation: every measured inference leaves
multiple recoverable copies in process memory. The small variation in
copy count does not change the underlying observation that plaintext
recovery occurs consistently after inference. Second, the sequential-tenant experiment shows that this residue is not
confined to the most recent request. Canaries associated with earlier
tenants remain recoverable after subsequent requests have completed.
Prompt lifetime therefore extends beyond both response generation and
the execution of later inference requests within the same serving
process. These observations expose a mismatch between application-level and
memory-level lifetime. From the application's perspective, a request has
completed and its temporary objects are no longer required. From a
forensic perspective, however, plaintext derived from that request
remains recoverable from the process address space.


{\bf Validating Runtime Residue.} We perform several checks to establish that the recovered plaintext
originates from the serving runtime rather than from LLAnalyzer's
acquisition procedure. First, prompt instances are tracked using fresh UUID canaries, enabling
recovered strings to be attributed to specific inference requests
through exact matching. Second, recovery is reproducible across
independent executions: although the precise number of copies may vary
slightly, residual prompt plaintext is recovered consistently after
inference. Third, the same behaviour is observed across multiple model
families and independent inference sessions, reducing the likelihood
that the observation is specific to a particular model or workload. Most importantly, the sequential-tenant experiment provides temporal
provenance for the recovered residue. Canaries associated with earlier
tenants remain recoverable after subsequent inference requests have
completed, demonstrating that the recovered plaintext represents
historical runtime state rather than merely the currently executing
request. Together, these checks support attribution of the recovered plaintext
to post-inference state retained within the serving process rather than
to content introduced by LLAnalyzer's acquisition procedure.


{\bf Takeaway.}
Prompt lifetime extends beyond inference completion in the evaluated
serving environment. Multiple plaintext representations remain
recoverable after response generation, and historical prompt contents
survive subsequent inference requests within the same serving process.
Runtime confidentiality therefore cannot be inferred from
application-level request completion alone.

We next investigate where these residual copies originate and why their
contents remain recoverable.

\subsubsection{RQ2.2: Why Does Prompt Residue Persist?}
\label{sec:rq2_mechanism}

RQ2.1 established that prompt plaintext remains recoverable after
inference and can persist across successive tenant requests within the
same serving process. We next investigate the mechanisms responsible
for extending prompt lifetime beyond the visible execution of an
inference request. Unlike conventional memory-forensics studies that identify surviving
objects without necessarily explaining their origin, LLAnalyzer
reconstructs the provenance of UUID-tagged prompt copies through
debugger-assisted runtime inspection. Rather than treating each
recovered copy as an independent artefact, we trace how a single user
prompt propagates through successive software layers during normal
inference execution.


{\bf Evidence.}
Debugger-assisted provenance analysis follows the prompt throughout the
inference pipeline, including HTTP request decoding, JSON parsing,
chat-template rendering, inference request construction, token
generation, and response teardown. Across the evaluated executions,
recovered UUID-tagged prompt data are associated with three recurring
mechanisms:
\textit{(1). Dynamic \texttt{std::string} growth.}
During prompt construction and template expansion, \texttt{std::string}
objects may repeatedly increase their capacity. When reallocation
occurs, the previous character buffer is released back to the allocator
without necessarily being overwritten immediately, allowing its
previous plaintext contents to remain recoverable.
\textit{(2). Transient runtime objects.} 
Intermediate processing stages, including JSON parsing, chat-template
construction, request serialisation, and related request-processing
operations, generate short-lived runtime objects containing copies or
derived representations of the original prompt. Although these objects
are destroyed after their immediate use, their underlying memory can
continue to contain recoverable prompt data. 
\textit{(3). Allocator-managed memory reuse.}
After object destruction, released buffers enter allocator-managed
reuse mechanisms, including per-thread caches and arena-managed regions
~\cite{glibc_arena}. These mechanisms retain released memory for subsequent
allocation rather than immediately clearing its previous contents,
thereby extending the period during which prompt plaintext can remain
recoverable. 
%
%
Collectively, these mechanisms account for the provenance of the
multiple prompt representations observed in RQ2.1.


{\bf Analysis.}
The provenance analysis shows that prompt persistence is not caused by a
single programming error or forgotten deallocation. Instead, it emerges
from the interaction of software components that create multiple
plaintext representations during ordinary inference.

Importantly, the analysis separates prompt \emph{duplication} from
\emph{retention}. Multiple independent representations are created
before memory enters glibc's reuse structures through JSON parsing,
chat-template expansion, request construction, serialisation, and
repeated \texttt{std::string} reallocations. After the corresponding
objects are released, allocator behaviour extends the period during
which their underlying contents remain recoverable. Application-level data handling therefore explains why multiple prompt
representations exist, whereas allocator-managed memory reuse explains
why released representations may survive beyond their intended
application lifetime. The distinction is important because these two
mechanisms imply different mitigation requirements.

If allocator retention alone were responsible for the observed residue,
clearing allocator-managed memory would be sufficient to restore
confidentiality. The debugger traces instead show that sensitive
plaintext is propagated through several stages of the inference
pipeline before those allocations are released. Effective sanitisation
must therefore account for the lifecycle of prompt-derived
representations across the serving stack rather than targeting only
allocator state.


{\bf Root-Cause Attribution.}
The debugger traces identify two successive mechanisms:
application-level prompt duplication followed by allocator-level
retention. Application processing creates the plaintext
representations; allocator behaviour subsequently extends their
recoverable lifetime. We therefore attribute prompt proliferation to
application-level data handling and persistence to subsequent memory
retention.


{\bf Takeaway.}
Prompt residue results from the combined effect of application-level
duplication and subsequent memory retention. Prompt data are copied
across multiple stages of request processing, while allocator behaviour
allows some of those representations to remain recoverable after their
application objects have been released. Runtime prompt confidentiality
is therefore a lifecycle problem spanning multiple software components,
rather than an isolated property of the memory allocator.

\textit{Next}, we investigate whether practical runtime mitigations can interrupt
this propagation chain and restore prompt confidentiality after
inference.

\subsubsection{RQ2.3: Can Runtime Sanitisation Restore Confidentiality at
Acceptable Cost?}
\label{sec:rq2_mitigation}

We now investigate whether these
runtime confidentiality failures can be mitigated in practice. Specifically, we ask two complementary questions:
\emph{(i) can runtime sanitisation meaningfully reduce residual prompt
exposure, and (ii) what performance cost does such protection impose on
normal inference?} A practical mitigation should reduce recoverable sensitive state without
materially degrading the interactive inference performance expected from
consumer local-LLM deployments.


{\bf Evidence.}
To evaluate mitigation effectiveness, we compare the baseline
implementation with the patched implementation described in
Section~\ref{sec:methodology}. Guided by the provenance analysis in
RQ2.2, the patch sanitises selected prompt-related runtime structures and
zeroises KV-cache state after use.

Table~\ref{tab:blast} compares the resulting memory exposure after
twelve sequential tenants execute within the same serving process. In
the baseline configuration, prompt information remains recoverable from
11 of the 12 tenants, comprising 110 plaintext copies and approximately
10.8\,MB of recovered data. With sanitisation enabled, the recovered
corpus decreases to 78 copies and approximately 3.35\,MB. This
corresponds to reductions of approximately 29\% in recoverable copy
count and 69\% in recovered plaintext volume. Despite these reductions, the tenant-level outcome does not change: prompt information remains recoverable from 11 of the 12 tenants under
both configurations. We separately measure the performance cost of KV-cache zeroisation using
50 inference requests per model. Table~\ref{tab:overhead} reports mean
request latency for patched and unpatched executions. Across the four
evaluated model families, the difference ranges from $-0.016$\,s to
$+0.019$\,s per request, which is below 0.6\% of mean request latency.

\begin{table}[!t]
    \centering
    \caption{Runtime-memory exposure after twelve sequential tenants.
    Sanitisation reduces recoverable prompt copies and plaintext volume,
    but prompt information remains recoverable from 11 of 12 tenants.}
    \label{tab:blast}
    \footnotesize
    \begin{tabular*}{\columnwidth}{@{}l@{\extracolsep{\fill}}lccr@{}}
        \toprule
        Configuration & Tenants & Copies & Plaintext (KB) \\
        \midrule
        Unpatched & 11/12 & 110 & 10\,834 \\
        Sanitised & 11/12 & \textbf{78} & \textbf{3\,352} \\
        \bottomrule
    \end{tabular*}
\end{table}

\begin{table}[!t]
    \centering
    \caption{KV-cache zeroisation overhead ($n=50$ requests per model).
    Mean request latency is reported in seconds.}
    \label{tab:overhead}
    \footnotesize
    \begin{tabular}{@{}lrrr@{}}
        \toprule
        Model & Patched & Unpatched & $\Delta$ (s) \\
        \midrule
        Nemotron-3-Nano-4B      & 2.822 & 2.838 & $-0.016$ \\
        Qwen3.5-9B              & 4.705 & 4.692 & $+0.013$ \\
        Gemma-4-E4B-it          & 2.908 & 2.891 & $+0.017$ \\
        Phi-4-reasoning-plus    & 8.422 & 8.403 & $+0.019$ \\
        \bottomrule
    \end{tabular}
\end{table}


{\bf Analysis.}
The mitigation results reveal an important distinction between reducing
\emph{residual exposure} and restoring \emph{confidentiality}.
Sanitisation removes approximately 29\% of recoverable prompt copies and
69\% of recovered plaintext volume, demonstrating that a substantial
fraction of residual sensitive state can be removed. However, prompts
from 11 of 12 previous tenants remain recoverable. The measured
tenant-level confidentiality outcome therefore remains unchanged. This result follows directly from the provenance analysis in RQ2.2.
Prompt plaintext propagates through several independently managed
runtime representations. Sanitising some of these representations can
substantially reduce the amount of recoverable information, but
confidentiality is not restored while another representation containing
the same tenant's prompt remains accessible.

Our performance measurements address a different aspect of the
mitigation. KV-cache zeroisation changes mean request latency by at most
20\,ms in our measurements, with differences below 0.6\%
across all four models. The small negative difference observed for
Nemotron ($-0.016$\,s) should not be interpreted as a performance
improvement; rather, the measured differences are sufficiently small
that the evaluated zeroisation operation introduces no practically
meaningful latency penalty under these experimental conditions. Together, these results suggest that the principal difficulty is not the
cost of clearing a known sensitive structure, but achieving sufficient
coverage over all prompt-bearing representations created during the
request lifecycle.


{\bf Residual Exposure Analysis.}
The remaining exposure provides evidence about where sanitisation must
operate. RQ2.2 showed that prompt-derived plaintext can appear during
template expansion, JSON processing, request serialisation,
\texttt{std::string} growth, and subsequent allocator-managed retention.
A mitigation that clears only selected structures therefore protects
only the copies passing through those sanitised paths. This distinction also affects how mitigation effectiveness should be
measured. Copy count and recovered byte volume quantify the
\emph{magnitude} of residual exposure, whereas the number of tenants
whose prompts remain recoverable captures whether historical
confidentiality has been restored. Under the latter criterion, reducing
110 copies to 78 and approximately 10.8\,MB to 3.35\,MB is beneficial,
but insufficient: the same 11 historical tenants remain exposed. Restoring the evaluated lifetime boundary therefore requires
lifecycle-wide handling of prompt-bearing representations, such that no
recoverable copy remains after its intended use. The results do not show
that runtime sanitisation is inherently incapable of achieving this
goal; rather, they show that the evaluated sanitisation coverage does
not yet encompass every surviving prompt representation.


{\bf Takeaway.}
Runtime sanitisation substantially reduces residual prompt exposure at
low measured performance cost, but it does not restore confidentiality
under the evaluated configuration. Recoverable copy count falls by
approximately 29\% and plaintext volume by 69\%, while KV-cache
zeroisation introduces less than 0.6\% latency difference.
Nevertheless, prompts from 11 of 12 previous tenants remain recoverable. The limiting factor in our experiments is therefore sanitisation
\emph{coverage}, not the measured cost of zeroising known sensitive
state. Restoring runtime confidentiality requires prompt-bearing data to
be identified and cleared across its complete software lifecycle.

We next examine whether these runtime confidentiality properties change
when model execution is accelerated through GPU offloading.

\subsubsection{RQ2.4: Does GPU Execution Fundamentally Change Runtime
Confidentiality?}
\label{sec:rq2_gpu}

RQ2.1 showed that prompt plaintext remains recoverable following
CPU-based inference, while RQ2.2 identified application-level duplication
and subsequent memory retention as the mechanisms extending prompt
lifetime. RQ2.3 further showed that runtime sanitisation can substantially
reduce residual exposure without restoring confidentiality under the
evaluated configuration. We now examine whether GPU-assisted inference
changes these runtime confidentiality properties.

A plausible hypothesis is that moving model computation to the GPU also
moves sensitive prompt state away from host memory, thereby reducing or
eliminating the residue observed during CPU inference. If so, the
lifetime-boundary failures identified above might primarily characterize
CPU-based execution rather than GPU-accelerated deployments.


{\bf Evidence.}
We repeat the runtime forensic experiment using Qwen3.5-9B with partial
CUDA offloading while preserving the prompt construction, UUID canaries,
inference parameters, and forensic procedure used in the CPU experiments. Table~\ref{tab:vram} summarises the results. Immediately after an
inference request, LLAnalyzer acquires a host-memory snapshot and
recovers nine UUID-tagged prompt canaries from CPU-resident host memory
despite GPU offloading. After two additional requests execute within the
same serving process, all sixteen injected canaries are recoverable,
showing that historical prompt data remain present across successive
GPU-assisted requests. In a separate idle-time measurement taken after inference, the same nine
canaries remain recoverable after 10\,s without additional requests.
Thus, the observed host-memory residue does not disappear merely because
GPU execution has completed or the server becomes temporarily idle.
Following termination of the serving process, no canaries are recoverable
from the process-associated memory regions examined by LLAnalyzer.

\begin{table}[!t]
    \centering
    \caption{\small Host-memory persistence during GPU-assisted inference
    using Qwen3.5-9B with partial CUDA offloading. The 10\,s idle
    observation is measured independently after the initial request.}
    \label{tab:vram}
    \footnotesize
    \begin{tabular}{@{}p{0.35\columnwidth}cp{0.36\columnwidth}@{}}
        \toprule
        Observation point & Canary hits & Interpretation \\
        \midrule
        After one request
            & 9
            & Immediate post-inference residue \\
        After two additional requests
            & 16
            & Historical canaries remain recoverable \\
        After 10\,s idle
            & 9
            & Initial residue remains recoverable \\
        After server termination
            & 0
            & No hits in examined process regions \\
        \bottomrule
    \end{tabular}
    \vspace{-0.5cm}
\end{table}


{\bf Analysis.}
The experiment shows that partial GPU offloading changes where model
computation occurs without eliminating the host-side prompt lifecycle.
Although portions of neural-network execution are transferred to the
GPU, prompt construction, JSON processing, chat-template expansion,
tokenisation, request construction, and host--device coordination still
involve host-side software. Prompt-derived plaintext can therefore exist
in CPU-managed memory before, during, and after GPU execution. The sequential-request experiment provides further evidence that GPU
offloading does not eliminate historical host-memory residue in the
evaluated configuration. After two additional requests, all sixteen
injected canaries are recoverable from host memory. The relevant
confidentiality issue is therefore not simply whether tensors or
attention computation execute on the CPU or GPU, but whether
prompt-bearing host-side representations are removed once their intended
use has ended. The idle-time experiment further separates persistence from active
computation. The same nine canaries remain recoverable after 10\,s
without additional inference activity, showing that request completion
and temporary inactivity do not automatically remove the observed
plaintext. Finally, the absence of recoverable canaries from the examined
process-associated regions after server termination bounds the lifetime
observed by LLAnalyzer to the active serving process. This result should
not be interpreted as evidence of secure physical-memory erasure:
process termination removes the mappings examined by our acquisition
procedure, but our experiment does not establish whether the underlying
physical pages are immediately overwritten.


{\bf Host-Side Attribution.}
The provenance results from RQ2.2 help explain why GPU offloading does
not remove the observed residue. The recovered prompt representations
originate from host-side stages such as JSON parsing, template expansion,
request construction, tokenisation, \texttt{std::string} operations, and
subsequent allocator-managed retention. These operations occur before or
around GPU execution and therefore remain relevant even when substantial
model computation is offloaded. This attribution also defines the scope of our result. Our experiment
directly evaluates Qwen3.5-9B under partial CUDA offloading; it does not
establish that every fully offloaded model or alternative GPU-serving
architecture exhibits identical persistence. Nevertheless, within the
evaluated serving architecture, relocating model computation does not
remove the host-side software paths responsible for the recovered
plaintext. GPU offloading therefore cannot substitute for explicit
lifecycle management of prompt-bearing host memory in this configuration.


{\bf Takeaway.}
GPU-assisted execution does not eliminate host-memory prompt residue in
the evaluated configuration. With Qwen3.5-9B under partial CUDA
offloading, prompt canaries remain recoverable after inference, survive
subsequent requests, and persist during an idle interval. These
observations are consistent with RQ2.2: sensitive prompt representations
are created and retained by host-side software independently of where
the model's computationally intensive operations execute. Accordingly, GPU acceleration should not itself be treated as a
host-memory confidentiality mechanism. Our measurements establish this
result for partial CUDA offloading; determining whether fully offloaded
models and other GPU-serving architectures exhibit the same lifetime
properties requires separate evaluation.

\subsection{RQ3. Persistence Boundary}
\label{sec:rq3}

RQ2 showed that prompt plaintext can outlive an inference request within
process memory. We next examine a distinct confidentiality boundary:
whether consumer wrapper applications independently extend prompt
lifetime by writing prompt-derived data to persistent storage. This distinction matters because runtime-memory sanitisation cannot
protect prompt data that have already been copied to wrapper-managed
files. We therefore ask: \emph{Do consumer local-LLM wrappers persist prompt plaintext beyond the
runtime lifetime of an inference request, and can such persistence be
disabled?}


{\bf Evidence.}
We evaluate LM~Studio and Ollama under their tested default
configurations by submitting UUID-tagged prompts and subsequently
searching the wrapper-managed files, logs, and configuration stores
included in our acquisition procedure for exact canary matches. The two wrappers exhibit different persistence behaviour. Under the
evaluated LM~Studio default configuration, UUID-tagged prompt plaintext
is recovered from a wrapper-managed plaintext log. Inspection of the
configuration identifies \texttt{logSensitiveData} as enabled in the
tested installation. Canary-tagged prompts from earlier experimental
sessions remain recoverable from these files days after the originating
inference requests have completed. We then disable \texttt{logSensitiveData} through
\texttt{http-server-config.json} and restart the application so that the
configuration change takes effect. Repeating the same canary-based
experiment yields no prompt-canary matches in the wrapper-managed
persistent artifacts examined by LLAnalyzer.

Ollama exhibits different behaviour under the evaluated configuration.
Across the wrapper-managed persistent artifacts included in our
acquisition procedure, LLAnalyzer recovers no UUID-tagged prompt
plaintext. As a complementary check for network-mediated disclosure, we execute
205 Ollama inference sessions inside an isolated network namespace while
monitoring outbound network activity. We observe no external connection
attempts attributable to the inference sessions. The corresponding
LM~Studio experiment likewise produces no observed outbound prompt
transmission under the tested configuration.


{\bf Analysis.}
These measurements show that persistent prompt lifetime is determined
independently of the runtime-memory behaviour examined in RQ2. In the
evaluated LM~Studio configuration, an inference request can disappear
from active application state while a plaintext representation remains
available in wrapper-managed storage. Disk persistence therefore creates
a second lifetime for the same sensitive information, independent of
whether process memory is subsequently sanitised. The configuration experiment further localises this behaviour.
Disabling \texttt{logSensitiveData} removes the observed persistent
canary recovery from the artifacts examined by LLAnalyzer without
changing the underlying model or inference engine. The measured
persistence is therefore attributable to wrapper-level logging in this
configuration rather than being an unavoidable consequence of local
inference.

The contrast with Ollama reinforces this distinction. We observe no
corresponding persistent prompt copies in the Ollama artifacts examined
under the tested configuration. Thus, two wrappers executing local LLMs
can expose different persistence properties even though both keep model
inference on the user's device. Finally, the network-isolation experiments separate local persistence
from external transmission. We observe no outbound connections during
the monitored inference sessions. The confidentiality issue identified
here is therefore local persistence within the evaluated wrapper
configuration, rather than evidence that prompts were transmitted to an
external inference service.


{\bf Configuration Attribution.}
The LM~Studio intervention provides a direct configuration-level
comparison. With sensitive-data logging enabled, injected canaries remain
recoverable from wrapper-managed persistent storage; after disabling the
corresponding setting and restarting the application, the same
acquisition procedure yields no canary matches in the examined
persistent artifacts. This result bounds our conclusion to the wrapper versions,
configurations, and artifacts evaluated in this study. It does not imply
that LM~Studio can never create other prompt-bearing artifacts, nor that
Ollama never persists prompt information under every possible
configuration. Rather, it demonstrates that wrapper-level configuration
can independently determine whether prompt plaintext survives on disk
after inference.


{\bf Takeaway.}
Prompt persistence is not determined solely by the inference runtime.
Under the evaluated LM~Studio default configuration, prompt plaintext is
written to wrapper-managed persistent storage and remains recoverable
after the originating request has completed. Disabling sensitive-data
logging eliminates the observed persistence from the artifacts examined
by LLAnalyzer. Under the evaluated Ollama configuration, we observe no
corresponding persistent prompt copies. The Persistence Boundary is therefore independently controlled by
wrapper-level data-handling decisions. Protecting runtime memory alone
is insufficient if another software layer retains the same prompt on
persistent storage.

\subsection{RQ4. Isolation Boundary}
\label{sec:rq4}

RQ2 and RQ3 examine how long prompt data survive within a single serving
environment. We now consider a different confidentiality property:
whether one authenticated client can learn information associated with
another client sharing the same serving process. This distinction separates \emph{authentication} from
\emph{isolation}. Establishing that a client possesses a valid API key
does not necessarily establish which server-side resources that client
is authorised to access. We therefore ask: \emph{Does the serving interface preserve prompt confidentiality between
authenticated clients sharing the same \texttt{llama-server} instance?} We evaluate three disclosure surfaces: direct access to saved slot state,
model-mediated disclosure after state restoration, and timing leakage
through shared prompt-prefix caching.


{\bf Cross-Tenant Slot Restoration.}
When \texttt{--slot-save-path} is enabled, the
\texttt{/slots/:id\_slot} interface exposes operations for saving and
restoring conversation state. In our experiments, requests to this
interface require a valid API key, but access to saved slot state is not
bound to the authenticated client that created it. We test this property using two independently authenticated clients.
Tenant~A creates and saves conversation state containing a UUID-tagged
secret. Tenant~B, using a different valid API key, subsequently requests
restoration of Tenant~A's saved slot state. Across 200 controlled trials
(50 per evaluated model family), the cross-tenant restoration succeeds
in 200/200 trials (Wilson 95\% CI: $[98.1\%, 100.0\%]$). The observed behaviour is consistent with a missing resource-level
authorisation check: authentication determines whether a request may
reach the interface, while the supplied slot identifier determines which
saved state is restored without an observed ownership check binding that
state to the requesting principal. We map this condition to CWE-862
(Missing Authorization) and CWE-639 (Authorization Bypass Through
User-Controlled Key)~\cite{cwe862,cwe639}.


{\bf Model-Mediated Disclosure.}
We next test whether restored state can produce application-visible
disclosure without directly inspecting the underlying saved
representation. After Tenant~B restores Tenant~A's state, we ask the
model to continue the restored conversation and test the generated
response for the victim's UUID-tagged secret. The secret is reproduced in 126 of 200 trials (63\%; Wilson 95\% CI: $[56.1\%, 69.4\%]$). The outcome varies
substantially across models: Gemma and Qwen reproduce the secret in
50/50 trials, Nemotron in 25/50, and Phi-4-reasoning-plus in 1/50. Importantly, the 63\% generation rate does not bound the underlying
authorisation failure: cross-tenant state restoration itself succeeds in
200/200 trials. These two counts refer to different channels and are
reported separately: the 200/200 figure is the memory-residue
measurement (restored slot state is present after restoration), whereas
the 126/200 figure is the output-continuity measurement (the generated
continuation contains the secret); state presence does not by itself
establish secret reproduction. Model continuation is therefore a secondary disclosure
channel whose observable success depends on subsequent generation
behaviour.


{\bf Shared-Prefix Timing Leakage.}
The third experiment does not require saved slot state. We evaluate
whether shared prompt-prefix caching exposes whether a candidate prefix
has previously been processed by the server. For each model, we measure time-to-first-token (TTFT) for 50 requests
whose prefixes correspond to cached state and 50 matched uncached
requests under loopback conditions, and we repeat the measurement over a
real wide-area path (a \texttt{llama-server} instance on a remote server
accessed from a residential client; mean RTT 78.4\,ms, baseline jitter
2.4\,ms). Table~\ref{tab:wan_timing} reports both conditions. Cached
requests exhibit substantially lower TTFT than their uncached
counterparts, and a threshold classifier separating the two achieves an
estimated AUC of 1.000 for each of the four evaluated models under
loopback and WAN conditions alike.

We additionally conduct a Monte Carlo sensitivity sweep that resamples
the measured TTFT distributions with 10{,}000 trials per model at
increasing levels of Gaussian network jitter. AUC remains 1.000 for all
four models up to $\sigma=100$\,ms (broadband WAN scale). At
$\sigma=500$\,ms (severe-congestion scale) it declines to 0.8534
(Gemma), 0.9328 (Nemotron), 0.9114 (Qwen), and 0.9999 (Phi-4), and at
$\sigma=1500$\,ms Phi-4 remains distinguishable at 0.8968 while the
three smaller models degrade to 0.63--0.68. Thus, even without
recovering prompt plaintext directly, an observer capable of issuing
appropriately constructed requests and measuring response timing can
distinguish whether candidate prefix state is present in the shared
cache under the evaluated configuration, and the oracle tolerates
jitter well beyond the conditions of our loopback experiments.

\begin{table}[!t]
    \centering
    \caption{Timing-oracle robustness. Time-to-first-token (TTFT) for 50
    cached and 50 uncached requests per model, measured on loopback and
    over a real wide-area path (mean RTT 78.4\,ms, baseline jitter
    2.4\,ms). AUC is computed per model and per condition.}
    \label{tab:wan_timing}
    \scriptsize
    \setlength{\tabcolsep}{1.2pt}
    \resizebox{\columnwidth}{!}{%
    \begin{tabular}{@{}lccccc@{}}
        \toprule
        Model & $n$ (c/u) & TTFT cached (ms) & TTFT uncached (ms) & AUC (loopback) & AUC (WAN) \\
        \midrule
        Gemma-4-E4B-it       & 50/50 & 318 & 2998 & 1.000 & 1.000 \\
        Nemotron-3-Nano-4B   & 50/50 & 20  & 786  & 1.000 & 1.000 \\
        Qwen3.5-9B           & 50/50 & 47  & 1150 & 1.000 & 1.000 \\
        Phi-4-reasoning-plus & 50/50 & 35  & 980  & 1.000 & 1.000 \\
        \bottomrule
    \end{tabular}%
    }
\end{table}


{\bf Analysis.}
The three experiments expose different aspects of the same isolation
problem. The slot-restoration experiment provides the clearest violation. Two
clients can both be correctly authenticated while remaining
insufficiently isolated from one another's server-side state. The
200/200 restoration result therefore demonstrates why authentication
alone is not equivalent to tenant authorisation: possession of a valid
credential establishes access to the service, but does not establish
ownership of the requested slot state.

Model-mediated continuation demonstrates a second consequence of the
same restored state. Once another tenant's conversation state becomes
available to the model, sensitive information can propagate into newly
generated output. The lower 126/200 reproduction rate should not be
interpreted as partial protection by the isolation boundary because the
underlying cross-tenant restoration has already succeeded in every
trial. Rather, generation introduces an additional model-dependent step
between unauthorised state access and observable plaintext disclosure.

The timing experiment reveals a separate form of cross-client
information exposure. Shared prefix caching need not reveal the complete
prompt to disclose information about another request. A measurable
difference between cached and uncached prefixes can instead act as a
membership oracle over candidate prompt prefixes. The AUC of 1.000 shows
complete separation in our measured samples under both loopback and
wide-area conditions and up to broadband WAN jitter; it should not be
extrapolated to networks, loads, or deployment configurations beyond the
jitter levels we swept, at which separation does degrade. Together, these results show that tenant confidentiality depends on more
than API authentication. Saved state requires resource-level
authorisation, while shared optimisation state requires isolation or
other controls if its presence is observable across clients.


{\bf Isolation Attribution.}
The direct-restoration experiment localises the primary failure to the
serving interface rather than to model behaviour. Cross-tenant
restoration succeeds before model generation is involved and does so
across all 200 trials. The subsequent 63\% secret-reproduction rate is
therefore not the cause of the isolation failure, but one consequence of
state that has already crossed the tenant boundary. The timing channel is logically independent of slot restoration. It
requires neither saved conversation state nor successful plaintext
generation; instead, it arises from shared prefix-cache state whose
presence affects externally observable request latency. Consequently,
removing the slot-restoration path would address the direct
authorisation failure but would not, by itself, eliminate cache-mediated
information leakage.


{\bf Takeaway.}
The evaluated serving interface does not preserve tenant confidentiality
across the tested shared-state mechanisms. Cross-tenant slot restoration
succeeds in 200/200 trials despite clients using distinct valid API keys (Wilson 95\% CI: $[98.1\%, 100.0\%]$), and restored state subsequently produces the victim's secret in 126/200 model continuations (Wilson 95\% CI: $[56.1\%, 69.4\%]$). Independently, shared prefix caching produces
complete cached-versus-uncached separation in our measured TTFT samples
(AUC\,=\,1.000) under both loopback and real wide-area conditions, and
the separation persists up to broadband WAN jitter. These findings expose two distinct requirements for multi-client
serving: \emph{resource-level authorisation} for persistent or restorable
conversation state, and \emph{isolation of shared optimisation state}
whose presence can otherwise become externally observable. Authentication
alone does not provide either property under the evaluated
configuration.

\vspace{-0.2cm}
\section{Related Work}
\label{sec:related}
\vspace{-0.2cm}
{\bf Local LLM Serving Security.}
Recent work has exposed memory-safety and logic vulnerabilities in local
LLM serving stacks, including heap and integer overflows, malicious model
parsing, template injection, and authorization flaws affecting
\texttt{llama.cpp}, Ollama, SGLang, and vLLM
~\cite{cve_vocab_overflow,cve_bleedingllama,cve_jinja2_ssti,
cve_sglang_5760,cve_vllm_62164,cve_databricks,
cve_int_overflow_bypass,cve_ggml_nbytes,cve_2024_37032,
cve_ollama_updater}.
Studies of multi-tenant LLM serving further examine isolation properties
of shared inference architectures
~\cite{multitenant_llm2026,llmserving_survey2024}, while forensic studies
show that local LLM applications can retain prompts and conversation
artifacts~\cite{murtuza2026,localnotprivate2026,xu2026mobilellmforensics}.
These works establish individual attack surfaces; we instead study how
prompt confidentiality composes across the local serving stack.

{\bf Prompt Confidentiality and Shared Inference.}
Prior work has demonstrated confidentiality risks from shared inference
state, particularly timing leakage through prompt-prefix and KV-cache
reuse~\cite{wu2025iknowwhatyouasked,safekv2025,prefixwall2026,
wu2025iknowwhatyouasked}, and reconstruction of prompts from cached
representations~\cite{shadowcache2026,optileak2026}.
These studies establish that performance optimisations can expose
cross-request information. Our scope is broader: we examine whether
prompt confidentiality survives across model admission, runtime memory,
persistent wrapper state, and multi-client serving interfaces.

{\bf Memory Persistence and Sanitisation.}
Memory forensics and secure-memory research have long shown that freeing
an object does not necessarily erase its contents and have proposed
explicit zeroisation and secure allocation techniques
~\cite{explicit_bzero,securestring2025,zeroonfree2024,
procfs_forensics,glibc_arena}.
Recent LLM-focused forensic work similarly recovers sensitive
conversation artifacts from local and edge deployments
~\cite{murtuza2026,xu2026mobilellmforensics}.
We build on these techniques to trace prompt copies through the inference
lifecycle and evaluate whether runtime sanitisation actually removes
cross-request exposure.

{\bf Prompt Injection and Agentic Security.}
Prompt injection targets a complementary layer of the LLM ecosystem.
Prior studies demonstrate direct and indirect instruction injection
against LLM-integrated applications and agentic systems
~\cite{greshake2023,perez2022,ferrag2025promptinjection,
agentprompt2026,kim2026agenticlandscape,agentsurvey2026}.
These attacks manipulate model or agent behaviour through adversarial
input; our attacks do not require the model to follow malicious
instructions. Instead, we study confidentiality properties of the
software infrastructure that processes and retains otherwise benign
prompts.

{\bf Positioning.}
Prior work therefore provides strong evidence for individual failure
modes---parser vulnerabilities, cache side channels, residual memory,
and prompt injection---but treats them largely as separate problems.
LLAnalyzer connects these surfaces through four explicit confidentiality
boundaries: \emph{Integrity}, \emph{Lifetime}, \emph{Persistence}, and
\emph{Isolation}. This framing allows us to determine not only whether
prompt information leaks, but \emph{where in the serving stack the
confidentiality guarantee fails}.
\vspace{-0.15cm}
 \section{Conclusion}\label{sec:conclusion}
 \vspace{-0.15cm}
In this paper, we introduced \textbf{LLAnalyzer}, a boundary-oriented framework
for evaluating prompt confidentiality across local LLM serving systems.
By decomposing the serving stack into four independently measurable
boundaries---\emph{Integrity}, \emph{Lifetime}, \emph{Persistence}, and
\emph{Isolation}---we show that local execution alone does not provide
end-to-end prompt confidentiality. 

Our measurements reveal sharply different properties across these
boundaries. We observed no parser bypasses or memory-safety violations
within the explored model-loading state space, whereas prompt plaintext
persisted in runtime memory beyond inference completion. Wrapper-level
persistence depended on configuration, while the serving interface
exposed the most consequential failures, including cross-tenant state
recovery and timing leakage from shared prompt-prefix caching measured
under both loopback and real wide-area conditions. Runtime
sanitisation substantially reduced residual plaintext at negligible
performance cost, but did not restore confidentiality because surviving
prompt representations remained recoverable. 

These results show that \emph{locality is a deployment property, not a
confidentiality guarantee}. Protecting prompts requires lifecycle-wide
controls over how sensitive data are admitted, copied, retained, and
shared across the serving stack. More broadly, LLAnalyzer provides a
methodology for evaluating these guarantees as local LLM systems evolve
across serving frameworks and hardware platforms.

\appendix
\bibliographystyle{plainurl}
\bibliography{references}

\appendix
\section*{Open Science}
\label{app:open_science}

We release LLAnalyzer together with the experimental harnesses,
configurations, analysis scripts, patches, and supporting artifacts used
to produce the results reported in this paper. Our goal is to make each
boundary-specific measurement independently reproducible rather than
provide only the final aggregate results. The artifact therefore contains
the experimental harnesses and configuration information required to
reproduce our model-loading, runtime-memory, wrapper-persistence, and
serving-interface experiments. For anonymous review, the artifact is
available at:
\url{https://github.com/0xzodiac/Local-LLMs-Forensics}.

Where licensing or redistribution restrictions prevent us from
redistributing third-party model artifacts, we instead provide the exact
model identifiers, versions, SHA-256 digests, and acquisition
instructions. These identifiers allow an independent evaluator to
reconstruct the model set used in our experiments and verify that the
resulting artifacts are byte-identical to those evaluated in this study.

A collaborator independently reran the framework on a separate research
server and reported matching results across all four boundaries,
including the timing boundary; the closure artifact is the per-request
dataset together with the original analysis script and manifest.

\section*{Ethical Considerations}
\label{app:ethics}

We designed all experiments to avoid exposing data belonging to real
users. We conducted the experiments exclusively on systems and accounts
under our control and used synthetic UUID-tagged prompts as experimental
secrets. We did not collect human-subject data, access third-party
conversations, or test cross-tenant attacks against production systems.

For the isolation-boundary experiments, we instantiated both the victim
and attacker tenants ourselves. We performed state-recovery and timing
experiments only against researcher-controlled serving instances and
never attempted to recover prompts or conversation state from publicly
accessible LLM servers.

Our Internet measurements were limited to passive service metadata and
non-destructive availability checks. We did not exploit, modify, or
retrieve state from discovered hosts. Where our experiments uncovered
security-relevant implementation behaviour, we followed coordinated
vulnerability-disclosure practices and withheld artifact details where
premature release could create unnecessary risk.

\input{files/appendix.tex}
\end{document}

%% file: files/appendix.tex
\section{Experimental Methodology}
\label{app:methodology}

We designed our experiments around a common question that applies to all
four boundaries: after a prompt crosses a particular software boundary,
can information attributable to that prompt still be recovered where the
corresponding confidentiality objective says it should no longer be
accessible?

To answer this question consistently, we use the same experimental
ground-truth strategy throughout the study. We embed fresh UUID canaries
in controlled prompts, record their association with the originating
request, execute the target operation, and then search the
boundary-specific evidence source for exact recovery. What changes
between experiments is therefore the boundary and evidence source, not
the basic criterion used to establish disclosure.

Unless otherwise stated, we repeat each model-dependent experimental
condition at least 50 times for each of the four evaluated model
families. We keep hardware, model artifacts, inference parameters, and
software revisions fixed except where varying one of these factors is
itself part of the experiment.

\subsection{Hardware and Software Configuration}
\label{app:hardware}

\paragraph{Primary test environment.}
We conduct the main experiments on a dedicated bare-metal Ubuntu
workstation and use the same machine for the model-file, runtime-memory,
wrapper-application, and serving-interface measurements. We avoid
unrelated user workloads during measurement runs to reduce interference
from background memory allocation, scheduling, and network activity.

Table~\ref{tab:hardware_config} records the hardware and operating-system
configuration used for the reported experiments.

\begin{table}[!b]
    \centering
    \caption{\small Hardware and operating-system configuration used in
    our experiments.}
    \label{tab:hardware_config}
    \footnotesize
    \setlength{\tabcolsep}{4pt}
    \begin{tabular}{ll}
        \toprule
        \textbf{Component} & \textbf{Configuration} \\
        \midrule
        Operating system &
        Ubuntu Linux, bare-metal 24.04.4 LTS \\
        Linux kernel &
        7.0.0-29-generic \\
        CPU &
        Intel Core i7-9850H CPU @ 2.60GHz \\
        System memory &
        16\,GB (15\,GiB reported) \\
        GPU &
        CUDA-capable GPU with 8\,GB VRAM \\
        GPU driver &
        NVIDIA 595.84 \\
        CUDA toolkit &
        12.0 (V12.0.140) \\
        File system &
        ext4 \\
        \bottomrule
    \end{tabular}
\end{table}

We run the GPU-memory experiments directly on the bare-metal Ubuntu host
rather than through a virtual machine or compatibility layer. This gives
us direct access to \texttt{cuda-gdb} and avoids introducing an
additional GPU-memory-management layer into the measurements.

\paragraph{Serving software.}
We use \texttt{llama.cpp} and its \texttt{llama-server} component for the
core runtime-memory and serving-interface experiments. For the
wrapper-level measurements, we additionally evaluate LM~Studio and
Ollama. We inspect selected components of the wider local-LLM ecosystem,
including GGUF Python tooling and SGLang, where required by the
boundary-specific experiments.

\begin{table}[!t]
    \centering
    \caption{Software components used in our evaluation.}
    \label{tab:software_config}
    \footnotesize
    \setlength{\tabcolsep}{4pt}
    \begin{tabular}{@{}%
        >{\raggedright\arraybackslash}p{\dimexpr0.25\columnwidth\relax}%
        >{\raggedright\arraybackslash}p{\dimexpr0.4125\columnwidth\relax}%
        >{\raggedright\arraybackslash}p{\dimexpr0.25\columnwidth\relax}@{}}
        \toprule
        \textbf{Component} &
        \textbf{Purpose} &
        \textbf{Version} \\
        \midrule
        \texttt{llama.cpp} &
        Runtime and serving-interface evaluation &
        Commit \texttt{388d39f3e} \\
        \texttt{llama-server} &
        Multi-slot and cross-tenant experiments &
        Same commit as \texttt{llama.cpp} \\
        LM~Studio &
        Wrapper logging and network behaviour &
        0.4.21 (Build 2) \\
        Ollama &
        Wrapper and network behaviour &
        0.20.5 \\
        SGLang &
        Reproduction check for GGUF/Jinja hardening &
        0.5.13.post1 \\
        glibc allocator &
        Heap-residue and allocator experiments &
        2.39 \\
        \texttt{gdb} &
        Host-memory inspection &
        15.1 \\
        \texttt{cuda-gdb} &
        GPU-memory inspection &
        12.0 (V12.0.140) \\
        Python &
        Measurement harnesses and analysis &
        3.14.6 \\
        \bottomrule
    \end{tabular}
\end{table}

\paragraph{Runtime configurations.}
We use the default \texttt{glibc} allocator configuration as our runtime
baseline. To determine whether allocator configuration changes the
observed prompt residue, we separately evaluate:

\begin{itemize}
    \item \texttt{MALLOC\_ARENA\_MAX=1}, which restricts the number of
    per-thread allocator arenas; and
    \item \texttt{MALLOC\_PERTURB\_=165}, which overwrites allocated or
    released memory with a deterministic byte pattern.
\end{itemize}

These controls require neither source-code modification nor binary
recompilation. We evaluate them separately from the code-level
zeroisation patch so that we can distinguish allocator-level effects
from explicit application-level sanitisation.

\paragraph{Isolation and network monitoring.}
We execute network-behaviour experiments inside an isolated network
namespace and monitor outbound connections throughout each run. We
exclude loopback communication between the wrapper and its local
inference runtime from the external-connection count.

\paragraph{Patched and unpatched binaries.}
For the runtime-mitigation experiments, we compare an unmodified baseline
against instrumented binaries implementing the sanitisation invariants
identified through our provenance analysis. The principal patch
configurations are:

\begin{description}
    \item[$I_2$:] deterministic sanitisation of released KV-cache cells;
    \item[$I_3$:] sanitisation of selected request and response buffers;
    and
    \item[$I_4$:] secure allocator handling for selected sensitive
    objects.
\end{description}

For each comparison, we keep the model, prompt sequence, inference
parameters, and runtime configuration identical between the patched and
unpatched conditions.

\subsection{Experimental Corpora, Models, and Parameters}
\label{app:models_parameters}

We do not use a conventional training or benchmark dataset because our
objective is not to measure model accuracy. Instead, we construct
boundary-specific experimental corpora that provide known ground truth
for each confidentiality test. These corpora comprise malformed GGUF
artifacts, synthetic UUID-tagged prompts, controlled multi-tenant
conversations, wrapper-persistence workloads, timing probes, and
Internet deployment metadata.

\paragraph{Evaluated models.}
We evaluate four model families selected to vary architecture, parameter
scale, chat-template behaviour, stopping behaviour, and reasoning style.
This diversity allows us to distinguish serving-stack behaviour that
persists across models from effects attributable to a particular model
family.

\begin{table}[!t]
    \centering
    \caption{Models used throughout the boundary-oriented
    evaluation.}
    \label{tab:evaluated_models}
    \footnotesize
    \setlength{\tabcolsep}{4pt}
    \scalebox{0.93}{\begin{tabular}{lll}
        \toprule
        \textbf{Model} &
        \textbf{Short name} &
        \textbf{Role in evaluation} \\
        \midrule
        \texttt{gemma-4-e4b-it} &
        Gemma &
        Instruction-following architecture \\
        \texttt{qwen-3.5-9b} &
        Qwen &
        Reasoning-capable model family \\
        \texttt{nemotron-3-nano-4b} &
        Nemotron &
        Compact reasoning model \\
        \texttt{phi-4-reasoning-plus} &
        Phi-4 &
        Explicit reasoning-output model \\
        \bottomrule
    \end{tabular}
}            
\end{table}

We execute all models locally using their GGUF-compatible
representations. Unless an experiment explicitly requires stochastic
generation, we use greedy decoding to reduce variation between repeated
trials.

\paragraph{Model-file corpus.}
We evaluate the integrity boundary using complementary structured and
coverage-guided approaches. The structured corpus contains malformed GGUF
artifacts targeting tensor dimensions, names, metadata lengths, header
fields, and related consistency constraints. We additionally exercise
the parser through coverage-guided fuzzing to explore states not captured
by manually constructed mutations.

For the structured component, we use:

\begin{enumerate}
    \item an 80-case malformed-input battery targeting the
    post-allocation-validation path of the C++ GGUF reader; and
    \item a 21-file structural-header corpus containing malformed tensor
    dimensions, tensor names, metadata lengths, and related header
    inconsistencies.
\end{enumerate}

We load each artifact independently and classify the resulting behaviour
as clean rejection, handled parsing error, unhandled exception, process
termination, successful malformed loading, or memory-safety failure. We
also process the 21 structural-header artifacts through the repository's
\texttt{gguf-py} tooling to compare behaviour between the C++ and Python
parsers.

In addition to these structured tests, we conduct the 24-hour
coverage-guided AFL++ campaign reported in the main paper, comprising
more than $1.2\times10^{7}$ parser executions under ASan and UBSan
instrumentation.

\paragraph{Synthetic confidentiality workload.}
We establish experimental ground truth using fresh UUID canaries
generated separately for each tenant and trial. Before injecting a
canary, we verify that it does not occur in the corresponding
pre-experiment evidence source. We then embed the identifier directly
into the controlled prompt and record its association with the
originating request.

This design gives us a direct attribution criterion. When we
subsequently recover an exact canary from process memory, a
wrapper-managed file, a protocol response, or another tenant's execution
context, we can associate that occurrence with the request in which we
introduced it rather than infer its provenance from surrounding
plaintext.

Depending on the confidentiality boundary under evaluation, we record a
direct disclosure when the originating request's canary appears in:

\begin{itemize}
    \item process heap memory;
    \item KV-cache or associated runtime buffers;
    \item GPU or pinned host memory;
    \item wrapper-generated files or logs;
    \item protocol response content;
    \item reasoning-output fields; or
    \item another tenant's continuation response.
\end{itemize}

\paragraph{Runtime-memory workload.}
For each model, we execute sequential multi-request sessions and inspect
memory after request completion and after logical release of the
associated session state. We repeat these measurements under:

\begin{enumerate}
    \item the unmodified runtime;
    \item each code-level mitigation configuration;
    \item \texttt{MALLOC\_ARENA\_MAX=1};
    \item \texttt{MALLOC\_PERTURB\_=165}; and
    \item selected combinations of code-level and allocator-level
    mitigations.
\end{enumerate}

We measure both the number of residual cleartext prompt copies and the
aggregate number of bytes associated with matching regions. For the
KV-cache sanitisation experiment, we additionally inspect all 56
targeted cache cells after release to verify whether the sanitisation
invariant holds at the individual-cell level.

\paragraph{Wrapper-application workload.}
We submit repeated UUID-tagged prompts to LM~Studio and Ollama and inspect
their application-managed files before and after each run. We evaluate
the default logging configuration and, where supported, repeat the
measurement after explicitly disabling sensitive-data logging.

For the network-isolation experiment, we execute 205 Ollama inference
runs while recording every attempted external connection. We treat
loopback communication required for local inference as expected internal
traffic rather than an external disclosure.

\paragraph{Cross-tenant serving workload.}
We configure \texttt{llama-server} with two logical tenants sharing a
single serving process. Both clients possess valid server credentials;
the security property under test is therefore \emph{authorisation}, not
authentication. A correctly isolated server should allow each client to
operate its own state without permitting either client to recover state
belonging to the other.

For each trial, we create a victim session containing a fresh
UUID-tagged secret, save the victim's slot state, and then attempt to
restore that state from a second authenticated client's context. We score
the trial as a direct disclosure when the second client successfully
obtains state originating from the victim.

We perform 200 direct slot-transfer trials in total, comprising 50 trials
for each model. Each trial therefore consists of:

\begin{enumerate}
    \item creating a victim session containing a unique secret;
    \item saving the victim's slot state;
    \item attempting to restore that state from the second client's
    context; and
    \item determining whether the second client can recover the victim's
    state or secret.
\end{enumerate}

We separately evaluate whether restored state can induce model-level
continuation of the victim's secret. The primary continuation experiment
uses greedy decoding. An additional probabilistic reconstruction
experiment records the top-20 token probabilities at each harvested
position to determine whether unsuccessful greedy reconstructions
nevertheless retain recoverable probability mass associated with the
target secret.

\paragraph{Timing workload.}
We evaluate whether shared prefix caching creates an externally
observable signal about another tenant's prior prompt processing. For
each model, we collect 50 cached and 50 uncached measurements under
otherwise identical serving conditions and use time-to-first-token
(TTFT) as the observable. TTFT isolates the computational effect of
prefix-cache reuse before full-response generation introduces additional
variation.

Our attacker possesses valid access to the shared service but cannot read
another tenant's plaintext prompt, process memory, or saved slot state.
The attacker's task is narrower: given a candidate prompt prefix,
determine whether that prefix has previously been processed by the
shared server. We quantify this distinguishability using ROC AUC rather
than selecting a threshold specific to a single experimental run.

\paragraph{Reasoning-channel workload.}
We separately inspect conventional response content and
reasoning-specific response fields when evaluating model-dependent
disclosure. We record a disclosure when the target secret appears in
either channel. We make this distinction because some evaluated models,
particularly Qwen and Phi-4, may place recovered information in
reasoning fields that are not exposed through the conventional response
content.

\paragraph{Internet deployment data.}
We measure Internet prevalence passively using search-engine results for
identifiable Ollama and \texttt{llama.cpp} services. We do not exploit,
modify, or conduct cross-tenant experiments against public hosts. We
derive counts from service banners and protocol-level fingerprints and
limit live-host validation to non-destructive availability checks.

\paragraph{Summary of experimental scales.}
Table~\ref{tab:experiment_scale} summarises the principal workloads used
throughout the study.

\begin{table}[!t]
    \centering
    \caption{Scale of the principal experimental workloads.}
    \label{tab:experiment_scale}
    \footnotesize
    \setlength{\tabcolsep}{4pt}
    \scalebox{0.95}{\begin{tabular}{lrl}
        \toprule
        \textbf{Experiment} &
        \textbf{Scale} &
        \textbf{Unit} \\
        \midrule
        Structured GGUF battery &
        80 &
        malformed inputs \\
        Structural-header analysis &
        21 &
        malformed files \\
        Coverage-guided fuzzing &
        $>1.2\times10^{7}$ &
        executions \\
        KV-cell sanitisation &
        56 &
        inspected cache cells \\
        Cross-tenant slot transfer &
        200 &
        trials; 50 per model \\
        KV continuation &
        200 &
        trials; 50 per model \\
        Probabilistic reconstruction &
        200 &
        trials; 50 per model \\
        Timing side channel &
        400 &
        observations; 100 per model \\
        Stealth KV backdoor &
        200 &
        trials; 50 per model \\
        Concurrent slot manipulation &
        200 &
        trials; 50 per model \\
        Ollama network monitoring &
        205 &
        inference runs \\
        \bottomrule
    \end{tabular}
}
\end{table}

\section{Statistical Analysis}
\label{app:statistics}

We choose the statistical treatment according to the observable produced
by each boundary experiment rather than applying a single statistical
test across all measurements. Our analysis emphasises repeated
measurement, effect magnitude, and explicit reporting of both positive
and negative results.

\paragraph{Binary outcomes.}
For experiments producing a success-or-failure outcome, we report the
observed proportion

\[
\hat{p}=\frac{x}{n},
\]

where \(x\) is the number of successful disclosures or anomalous trials
and \(n\) is the total number of trials.

We report binomial uncertainty using two-sided 95\% Wilson score
intervals~\cite{wilson1927}. We use Wilson intervals because several of
our measurements lie close to the boundaries of zero or one, where
normal approximations are inappropriate. For example, observing zero
successes in 50 trials does not establish a true probability of zero;
the corresponding Wilson interval instead provides an explicit bound on
the unobserved event probability.

\paragraph{Deterministic findings.}
We use the term \emph{deterministic} only when every repeated trial under
a fixed experimental condition produces the same outcome. Examples
include direct slot-state transfer succeeding in 200 of 200 trials and
sanitisation succeeding for all 56 inspected KV-cache cells. Our use of
the term refers to repeatability under the evaluated configuration and
does not imply universal behaviour across all software versions,
hardware platforms, or deployment environments.

\paragraph{Memory-residue measurements.}
For runtime-memory experiments, we report:

\begin{itemize}
    \item the number of residual cleartext copies;
    \item the aggregate bytes occupied by matching memory regions; and
    \item the relative reduction from the unmodified baseline.
\end{itemize}

For baseline measurement \(B\) and mitigated measurement \(M\), we
calculate the relative reduction as

\[
\Delta_{\%}
=
100 \times \frac{B-M}{B}.
\]

We retain both copy counts and byte volumes because either metric alone
can obscure changes in residual exposure. A mitigation may substantially
reduce the number of surviving objects while leaving large plaintext
regions intact, or reduce byte volume while still leaving enough copies
to violate the confidentiality objective.

\paragraph{Timing-side-channel analysis.}
We retain the cached and uncached TTFT distributions separately for each
model and quantify their distinguishability using the area under the
receiver operating characteristic curve (AUC). This avoids selecting a
classification threshold tailored to one particular experimental run.

An AUC of 0.5 represents chance-level discrimination, whereas an AUC of
1.0 represents complete separation of the observed distributions. Thus,
the AUC of 1.000 observed for all four models means that, within our
collected measurements, the cached and uncached TTFT observations were
completely separable.

Where space permits, we provide the underlying TTFT distributions and
receiver operating characteristic curves in addition to the summary
statistics reported in the main paper.

\paragraph{Cross-model reporting.}
We report results separately for Gemma, Qwen, Nemotron, and Phi-4 where
model-specific behaviour affects the outcome. We avoid pooling these
measurements into a single aggregate proportion because differences in
chat templates, stopping behaviour, reasoning channels, and allocation
patterns are themselves relevant to the confidentiality analysis.

We describe a result as applying across model families only when we
observe the same qualitative behaviour for all four evaluated models.
Where quantitative differences occur, we retain and discuss them rather
than obscuring them through pooled averages. Figure~\ref{fig:forest}
summarizes the observed success rates with Wilson 95\% confidence
intervals for each boundary experiment.

\paragraph{Analysis of null results.}
We retain negative results when they meaningfully constrain a plausible
security hypothesis rather than omitting experiments that fail to
produce an attack. Before interpreting an observation as a negative
result, we verify that:

\begin{enumerate}
    \item the relevant instrumentation and response fields are active;
    \item an appropriate positive-control condition produces the expected
    observable;
    \item the serving process remains operational after the trial; and
    \item the null outcome is reproducible under the stated experimental
    condition.
\end{enumerate}

We therefore report the probabilistic reconstruction,
behavioural-injection, and concurrency experiments as bounded negative
results where appropriate rather than treating the absence of a
successful attack as evidence that the corresponding mechanism is
universally secure.

\paragraph{Reproducibility.}
We associate every quantitative claim with the corresponding measurement
script, configuration, or raw experimental log. The harness generates
randomised identifiers and UUID canaries and records them alongside the
trial in which they are introduced. Our analysis scripts derive the
reported tables and figures directly from these recorded measurements,
reducing manual transcription and preserving the mapping between raw
evidence and the results reported in the paper.